\documentclass[sigconf, nonacm]{acmart}

\AtBeginDocument{%
  \providecommand\BibTeX{{%
    \normalfont B\kern-0.5em{\scshape i\kern-0.25em b}\kern-0.8em\TeX}}}

\usepackage[bottom]{footmisc} 
\usepackage{subcaption}
\usepackage{tikz}
\newcommand{\pie}[1]{%
\begin{tikzpicture}
 \draw (0,0) circle (0.7ex);\fill (0.7ex,0) arc (0:#1:0.7ex) -- (0,0) -- cycle;
\end{tikzpicture}%
}

\begin{document}

\title{On the Educational Impact of ChatGPT: Is Artificial Intelligence Ready to Obtain a University Degree?}

\author{Kamil Malinka}
\affiliation{%
  \institution{Brno University of Technology}
  \streetaddress{Božetěchova 2}
  \city{Brno}
  \country{Czech Republic}
}
\email{malinka@fit.vutbr.cz}
\author{Martin Perešíni}
\affiliation{%
  \institution{Brno University of Technology}
  \city{Brno}
  \country{Czech Republic} 
  }
\email{iperesini@fit.vut.cz}
\author{Anton Firc}
\affiliation{%
  \institution{Brno University of Technology}
  \city{Brno}
  \country{Czech Republic} 
  }
\email{ifirc@fit.vut.cz}
\author{Ondřej Hujňák}
\affiliation{%
  \institution{Brno University of Technology}
  \city{Brno}
  \country{Czech Republic} 
  }
\email{ihujnak@fit.vut.cz}
\author{Filip Januš}
\affiliation{%
  \institution{Brno University of Technology}
  \city{Brno}
  \country{Czech Republic} 
  }
\email{ijanus@fit.vut.cz}

\renewcommand{\shortauthors}{Malinka and Perešíni, et al.}

\begin{abstract}
    In late 2022, OpenAI released a new version of ChatGPT, a sophisticated natural language processing system capable of holding natural conversations while preserving and responding to the context of the discussion.
    ChatGPT has exceeded expectations in its abilities, leading to extensive considerations of its potential applications and misuse.
    In this work, we evaluate the influence of ChatGPT on university education, with a primary focus on computer security-oriented specialization.
    We gather data regarding the effectiveness and usability of this tool for completing exams, programming assignments, and term papers.
    We evaluate multiple levels of tool misuse, ranging from utilizing it as a consultant to simply copying its outputs.
    While we demonstrate how easily ChatGPT can be used to cheat, we also discuss the potentially significant benefits to the educational system.
    For instance, it might be used as an aid (assistant) to discuss problems encountered while solving an assignment or to speed up the learning process.
    Ultimately, we discuss how computer science higher education should adapt to tools like ChatGPT.
\end{abstract}

\keywords{Academic Education, ChatGPT, Artificial Intelligence, Virtual Assistant, Computer Security}

\maketitle

\section{Introduction}
The end of 2022 brought, not only in the media, a big hype caused by the launch of a new version of the ChatGPT chatbot.

\begin{figure}[b]
  \includegraphics[width=\linewidth]{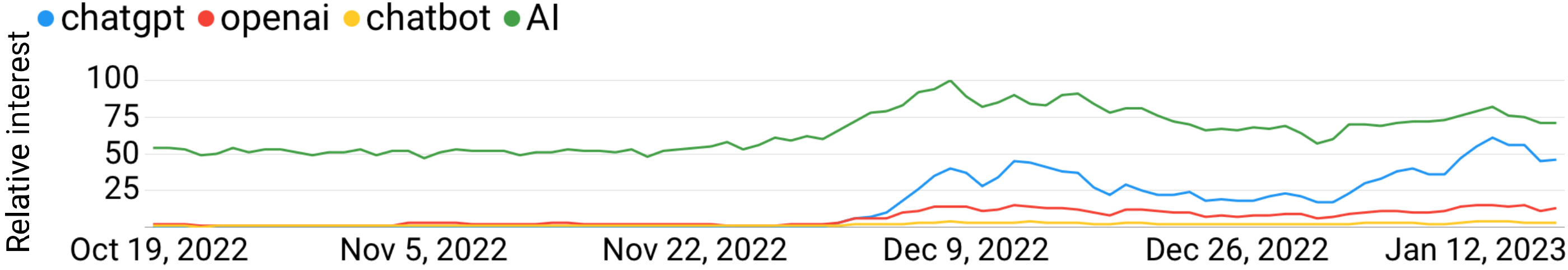}
  \caption{Trend in Google searches for chatbot and AI-related queries. Retrieved from Google Trends\protect \footnotemark.}
  \label{fig:trend}
\end{figure}

While the launch itself went unnoticed, astonished reactions from people playing with the technology soon flooded the media space. We can summarize the reactions as shocked, to say the least. Although the AI's capabilities were certainly not as shocking to the community dealing with this technology for a longer time, the availability to a broader audience caused a strong reaction. For example, this hype can be seen in the graph showing the search popularity for the relevant keywords (see \autoref{fig:trend}). Although it is not the first publicly available AI (e.g., DALL-E~\cite{dalle_2022} for image generation, FaceApp~\cite{faceapp} for face alterations, or Copilot~\cite{copilot} for code completion), it has certainly seen the biggest response as it gained over 1M users in under a week~\cite{mollman_2022}.

\footnotetext{Accessed 12.1.2023, \url{https://trends.google.com/trends/explore?date=today\%203-m&q=chatgpt,openai,chatbot,AI}}

Many people started to test the limits of the released software. These limits were found relatively early, yet the quality exceeded expectations. The tool is able to produce high-quality texts of various focuses even with the ability to respond in various languages (internally, they are machine translated into English similarly to DeepL~\cite{deepl} or Google Translate). Another strength of ChatGPT is contextual querying, where ChatGPT remembers previous queries and creates new results based on an earlier conversation.

Soon, signals emerged that this was the end of the classically written essays and, thus, the end of the teaching as we know it for selected disciplines. While this technology might be misused for cheating, there are also implications for technical fields, especially computer science, as it can simplify creative activities such as programming. A good example might be the GitHub Copilot~\cite{copilot} or already known first cases of malware created using ChatGPT~\cite{rees_2023}.

The student's ability to adopt this technology to subvert the text produced by ChapGPT proved to be very good and fast. Dutch students have already admitted to using this technology to write homework~\cite{nl_times_2023}. And as confirmed cases of cheating began to emerge very quickly, the reaction in education system was similarly swift. For example, New York City schools immediately banned this technology in the classrooms~\cite{rosenblatt_2023}. On the other hand, the first tools to detect ChatGPT-generated texts started to appear~\cite{syme_2023}. 

ChatGPT has also affected the computer security field and the phenomenon affected IT security in general.
Many colleagues have already experimented with solving their course exams or assignments. However, to our knowledge, there is no in-depth testing of ChatGPT capabilities. Another motivation is that our university already suspects that students are actively using AI tools to solve assignments. 

Our initial experiments have confirmed the great power of this technology. And not only in the English language. Most of the experiments were executed in the Czech language without difficulty, which, combined with real cases of student misuse, has led us to question whether we are far from AI (with the right user, of course) being able to complete standard college requirements successfully.

We, therefore, perform in-depth testing of ChatGPT capabilities on actual use cases, i.e., selected tasks that are representative of the standard requirements we place on students: exams, term papers, programming tasks, or the Capture The Flag exercise. We focus on the courses in the information security specialization.

For each evaluated task, we investigated different levels of AI assistance. From copying the results to using AI as an assistant to provide supplementary advice and information. For each task, we mark the AI-generated results using the same methodology as for regular students, and we compare them to average student results based on the statistics from previous years. 
In the scope of this paper, we abstract from the technical feasibility of the actual examination cheating. We are aware that the emphasis on supervision during the examination and the limitation of internet access is a solution, but not always, not everywhere (as we could see during the quarantine).

Based on the results, we describe the implications of this technology on traditional higher education. While the most discussed topics are usually malicious usage of AI (cheating, plagiarism) and it's effects (loss of critical thinking, etc.), we also see positive effects, which, with appropriate assistance, can significantly accelerate some phases of learning.
The ultimate goal of our papers is to open a discussion regarding the appropriate adoption of this technology in higher education. 

\subsection{Contributions}

The main contributions of this paper may be stated as follows:
\begin{itemize}
    \item We perform an in-depth evaluation of ChatGPT's abilities to solve assignments of various levels in computer security specialization at the IT-oriented university.
    \item We discuss this technology's possible positive and negative impacts on university education.
    \item Taking into account the capabilities of AI, we propose how the educational process (adaptation, detection, and prevention) should be revised to benefit both students and educators.
\end{itemize}

\subsection{Our position}
Higher education must respond adequately and immediately to the increasing quality of tools like ChatGPT. Their quality is already good enough to produce results comparable to the average student. An emphasis on modifying the examination process is inevitable.

Although we need to consider the power of the tools that students can use to cheat in their studies and adapt the environment to minimize these effects (pen-paper exams, explanation of negative effects, detection, etc.), we should not go down the path of mere prohibitions. And not just because of the lack of detection tools and, thus, the inability to punish misuse. Instead, we should find ways to prepare students for the effective use of these tools where they make sense - for example, in the form of an assistant with whom the student can \emph{"have a discussion"} and thus partially replace the peer (or teacher) in his absence. 

We are aware of the imperfections of the tool and that it can sometimes give wrong and misleading answers, but how is this different from discussions with classmates who have not yet mastered the content of the course? The quality of these tools will increase over time, and, in turn, their error rate will decrease. It should also be taken into account that these assistive tools allow for significantly faster learning on selected topics, which ultimately increases the overall effectiveness of learning. Moreover, their massive use in practice is to be expected, and students need to be prepared for these possibilities.

\vspace{-0.3cm}
\section{Related work}
\label{sec:relatedWork}
Researchers have been investigating the use of machine learning and artificial intelligence in education ever since these techniques reached sufficient maturity. The primary focus has been on improving the learning process by aiding the students and simplifying some repetitive processes for teachers.

Zong and Krishnamachari~\cite{Zong:MathWorldProblems:2022}, and Tack and Piech~\cite{Tack:AITeacher:2022} investigated the use of Natural language processing (NLP) for personalized teaching.
The first paper praises the ability of GPT-3 to extract correct equations from math world problems in 80\% of the tested tasks.
At the same time, the latter concludes that while conversational agents can communicate like a teacher and understand the student, they are lacking when it comes to helping the student. They both agree that applying common-sense knowledge remains an issue, and output can contain inconsistencies or random errors.

Moore et al.~\cite{Moore:QualityOfQuestionsByChatGPT:2022} tested the ability of GPT-3 to evaluate the quality of student-created test questions and classify them into Bloom's revised taxonomy, but their result did not achieve sufficient quality.

In his opinion paper on GPT-3~\cite{Dehouche:Plagiarism:2021}, Dehouche shows it can generate academic papers which pass current plagiarism checks, while Thunström and Steingrimsson ~\cite{GPT:CanGPTWriteAcademicPaper:2022} used it to generate a full scientific paper ( however, it contains typical flaws such as weak claims and unfit references). Cotton et al. ~\cite{Cotton:CheatingWithGPT:2023} demonstrated the use of ChatGPT as a powerful writing tool to generate a plausible paper.

The ability of GPT-3.5 to solve tests has been tested by Susnjak~\cite{Susnjak:ChatGPTEndOnlineExam:2022} and Bommarito and Katz~\cite{Bommarito:GPTBarExam:2022}. The latter applied the model to the multiple choice component of the law examination in the USA. They conclude that even though the examination requires understanding exam-specific questions with unique syntax, the model yields better results than expected, only 17\% worse than human-takers in general. Even though it would not manage to pass yet, it would surpass the passing range in some subcategories, which is in line with the results we obtained in our study.
However, in our case, the GPT-3.5 would succeed in some courses completely.

These possibilities raise the question about its ethical use, which is discussed by Tate et al.~\cite{Tate:EduResearchAIWriting:2023} who warn against banning its use but suggest their features shall be taught as well as their ethical usage (including the proper citation of AI usage). They argue that students must be taught to be cautious and challenge AI's responses.

\subsubsection{\textbf{Used Tools}}
This section describes the tools we used to complete assignments and detect AI-written text.    

\smallskip
\textbf{ChatGPT}~\cite{openai_2022} is a variant of the GPT (Generative Pre-trained Transformer) language model developed by OpenAI, specifically designed for natural conversation.
It is trained on a large dataset of human-generated conversations and can generate human-like responses to prompts given by a user. ChatGPT has been fine-tuned to generate responses that are more appropriate for conversational contexts and can generate coherent responses and maintain context across multiple turns of conversation. It is a powerful tool for generating human-like text in natural language conversation.

\smallskip
\textbf{GPTZero} is a ChatGPT-written text detector released shortly after the release of the chatbot~\cite{syme_2023}. It uses \textit{perplexity} and \textit{burstiness} metrics to differentiate between human and AI-written text. The human-written text normally has a higher perplexity -- the randomness of the text, and it exhibits properties of burstiness: non-common items appear in random clusters.

\section{Experiment design}
\label{sec:experimentDesign}
We have thoroughly analyzed the ability of AI to solve different levels of exams and projects of university courses on computer security. We let AI solve the tasks, assessed the results, and compared the received points with real students.
The testing sample may be improved; however, our goal is to show the basic concept and provide baseline data, not to cover all variants exhaustively.

We emphasize that Czech tests, assignments, etc., were used in our experiments, which may have caused differences compared to English assignments, but at the same time, broadened the scope and impact of the results.

First, we have established multiple exam categories we regularly use (see \autoref{matrix_categories}): written exams (fulltext questions, multiple-choice tests), term essays, and programming assignments (with designing and implementing source code and analytical tasks such as Capture the Flag~\cite{ctf}).
Each category is better described in the subsequent section.

Next, we set three modes of usage based on students' required expertise: naive AI usage -- cut and paste (low knowledge), minor clarification of answers -- interpretation of AI's answers (basic knowledge), and usage of AI as an assistant (good knowledge).
The matrix of different usages of AI and different categories are depicted in \autoref{matrix_categories}.
Finally, we select the number of real assignments from each category. We use ChatGPT to solve them in defined usage modes, evaluate them using the same methodology used for regular courses, and compare results.

\subsection{Categories of examination methods}

 For the experiment, we include examination methods from four courses focused on the security of information systems, cryptography, secure coding, and secure hardware. We define three basic types of methods: written exams, term essays, and programming assignments.
 
\subsubsection{\textbf{Written exams}}
Written exams are used to test a student's knowledge and his ability to apply it.  All exams combine general knowledge and practical questions (e.g., knowledge and understanding of key terms, encryption using transposition cipher, calculating the output of several runs of a specific LFSR generator, explanation of the principle of the selected attack, knowledge of the responsibilities of the information security administrator role, designing security measures for a given scenario, etc.). The used questions cannot be published due to yearly reuse.
We used two different types of written exams:

\begin{table}[t]
    \centering
    \footnotesize
    \begin{tabular}{l|cccc}
    \toprule
     & Test & Fulltext & Essay & Programming \\
    \midrule
    Copy\&Paste & \pie{180} & \pie{360} & \pie{180} & \pie{180} \\
    Interpretation & \pie{180} & \pie{0} & \pie{360} & \pie{360} \\
    Assistant & \pie{0} & \pie{0} & \pie{180} & \pie{360} \\
    \bottomrule
    \end{tabular}
    \vspace{0.2cm}
    \caption{Matrix of exam categories and AI usage modes \\
        \textit{\small 
        The \protect\pie{0} / \protect\pie{360} symbol represents whether AI was used or not for a specific examination category.
        A \protect\pie{180} indicates that the use of AI depends on the task or question at hand
        (e.g. in an essay setup, AI may be used for assistance, but sometimes it's only needed for Copy\&Paste).
        }
    }
    \label{matrix_categories}
    \vspace{-1.0cm}
\end{table}

\smallskip
\textbf{Fulltext exam.}
Fulltext exams require students to answer a~question using their own words or to demonstrate a solution to a problem. 

\smallskip
\textbf{Test.}
Students choose from a predefined set of responses where one or more is correct. The points for a question are assigned according to how many of the correct responses were selected. If any incorrect response is selected, a minus point is awarded for that whole question.

\subsubsection{\textbf{Term essays}}
Term essays require the students to study security-related topics and write a short research paper (approx. 4~-~6~pages) about collected knowledge. It is possible to write a tutorial paper, perform a security analysis of a certain product, or execute a new exploit and document it.

\subsubsection{\textbf{Programming assignments}}

We select three types of programming assignments used by our faculty.

\smallskip
\textbf{Completing predefined code.}
An individual project, where a predefined structure in the form of methods, variables, or usage definition is provided. Students are typically assigned to implement selected methods. We selected a machine learning class homework, where students implement formulas for simple distributions and classifiers in Python. This type of assignment allowed for testing both Copy\&Paste and interpreted variants. 

\smallskip
\textbf{Small project.}
An individual project typically consists of less than 1,000 lines of code. We selected an RSA implementation in C++. The requirements are to generate the keys, encrypt, decrypt, and break messages with small key lengths. Additionally, it is not allowed to use special libraries for the generation of prime numbers or primality tests.

\smallskip
\textbf{Term project.}
A large programming project, primarily for teams of two to three people. We selected a term project from the Information Systems course, as the security-related courses do not have this type of project. 
The assignment is to implement a simple information system as a web application in PHP. The topic was \emph{Smart City}. This system should allow the citizens of a city to report issues and the city manager to manage these tickets.

\smallskip
\textbf{Interactive project.}
An individual interactive project requires the student to get familiar with an unknown environment (system) and complete tasks. One of the courses involves the Capture The Flag (CTF) task. Students are given cybersecurity-related challenges, such as encryption cracking, reverse engineering, web exploitation, etc. Users must solve these challenges to retrieve hidden messages in the system.
The assignment consisted of six secrets within a~virtual network environment of four servers. Tasks are designed for all students level from beginners to intermediate.

\vspace{-0.25cm}
\subsection{\textbf{AI usage modes}}
We used ChatGPT in mainly three distinct modes (see \autoref{matrix_categories}).
The modes are primarily based on students' required apriori expertise in a given topic:

\begin{itemize}
    \item \textbf{Copy\&Paste} mode, where the necessary knowledge is non-existent, meaning the person only copies the given question into AI and pastes the result into the original questionnaire.
    \item \textbf{Interpretation} mode.
    This mode is more elaborated from the user's perspective.
    The user attempts to comprehend the AI-provided response and, based on his understanding, alters the final response.
    The responses are usually verbose and occasionally contradictory, but with the user's awareness, he uses only valid portions or creates his answer for provided facts.
    \item \textbf{Assistant} mode.
    In the most advanced mode, the user must fully comprehend the AI outputs, correct any possible errors, and ask specific questions to gain the most value from the offered responses to maximize the AI system's potential.
    This option requires interaction with the AI and more time than other modes. 
\end{itemize}

\smallskip
\noindent
\textit{\textbf{Usage modes with examination categories:}}

\begin{itemize}
    \item \textbf{Fulltext exam.}
    For fulltext exams, the ChatGPT was used purely in Copy\&Paste mode.
    \item \textbf{Test.}
    As ChatGPT provides both the responses and explanations, it is possible to improve the results by interpreting them. We thus examine two scenarios, Copy\&Paste, and interpretation of responses.
    \item \textbf{Term essays.}
    Two levels of usage to write essays are available: \textit{completely written by AI}, \textit{help from AI}. In the second case, we replaced some parts of the student's paper with ChatGPT written text.
    \item \textbf{Programming assignments.}
    As in most cases, ChatGPT is used as an assistant; instead of implementing numerous assignments, we carefully examined the code generated by ChatGPT and its recommendations on how to solve the given problem. Using this knowledge and comparing it with the original student's work, we estimated the range of points gained by ChatGPT. We also evaluate the usability of ChatGPT to solve these CTF-based projects.
    Because of this setting, we evaluate only the overall usability instead of awarding points.
\end{itemize}

\subsection{\textbf{Experiment scope}}
Ultimately, we evaluated four courses related to computer security.
The courses consist of different combinations of examination methods and their weights for final assessment.

We used a minimum of 50 assignment types for each category of examination method (except essays, where there were created twenty different essays).
The assignments were solved by AI and compared with the results of students solving exams from the same question pool.

For CTF, we used an approach, where only one AI-based solution was relevant.
In the scope of programming assignments, a different number of tasks were used depending on difficulty, i.e., small project, term project, and code completion.

\section{Evaluation}
\label{sec:evaluation}
\begin{figure}[b]
    \includegraphics[width=\linewidth]{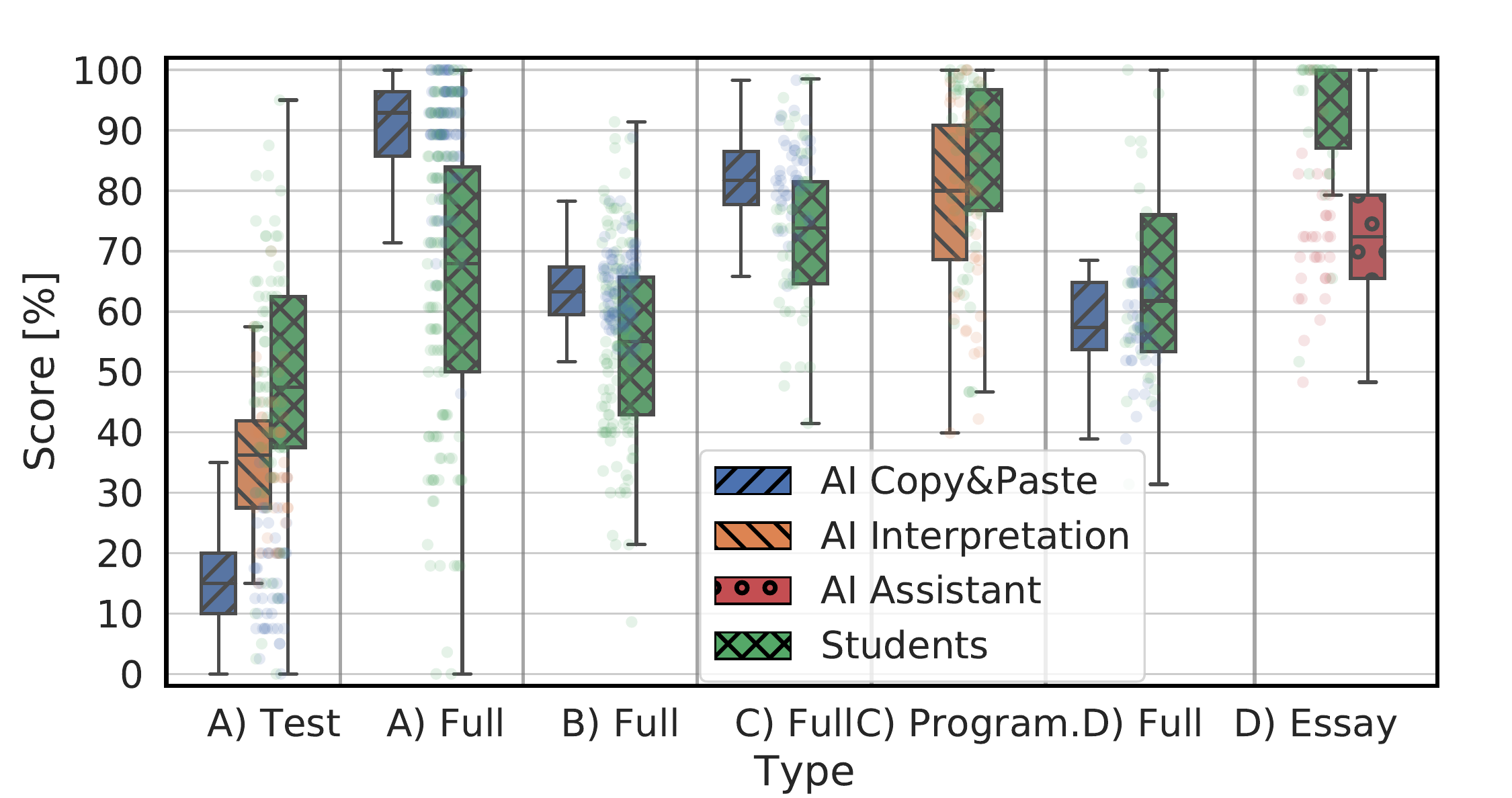}
    \caption{The results obtained by scoring AI outputs in multiple evaluation categories with comparison to a genuine sample of student scores.}
  \label{fig:evaluation}
\end{figure}

This section presents an evaluation and interpretation of our observations.
\autoref{fig:evaluation} displays all scoring of given varied tasks done by AI compared to genuine student performance.
Note that in this case, capital letters A)$\ldots$D) represent four examined courses.
We organize experiments using the task categorization proposed in \autoref{sec:experimentDesign}.

\smallskip
\noindent
\emph{\textbf{Fulltext exam.}}
\autoref{fig:evaluation} [\texttt{A) Full}, \texttt{B) Full}, \texttt{D) Full}] compares AI to students regarding points received from fulltext exam questions.
ChatGPT provides correct answers, but they are broader and less specific.
Problems can arise when answering questions requiring knowledge application, such as when employing the Bell-Lapadua model for document access; ChatGPT has no sense of context in this situation.
The variability of answers is similar to ordinary students.

\smallskip
\noindent
\emph{\textbf{Tests.}}
\autoref{fig:evaluation} [\texttt{A) Test}] demonstrates that the selected solutions are not always correct in the \textit{Copy\&Paste} mode.
This is deteriorated by the fact that this specific test has harsh scoring, with an incorrect answer resulting in a -1 score for a given question.
\textit{Interpretation} mode, on the other hand, considerably improves response quality, as shown in the graph.

\smallskip
\noindent
\emph{\textbf{Term essays.}}
\autoref{fig:evaluation} [\texttt{D) Essay}] shows that AI performed worse than students.
The approximately four-page Copy\&Pasted essay was completed in less than an hour; only with assistance did it take longer, as fact-checking was required.
However, the generation of, for example, threat modeling instrument descriptions saved time.

\smallskip
\noindent
\emph{\textbf{Completing Predefined Code.}}
ChatGPT allowed us to complete the homework without looking up all the needed formulas or algorithm steps (e.g., Expectation-Maximization training for Gaussian Mixture Model).
Copy-pasting the predefined structure to ChatGPT allowed easy code generation with needed properties for the template.
Even a naive implementation (\emph{copy\&paste}) received at least 30\% of points with no more than 10 minutes of spent time.

\begin{figure}[t]
  \begin{subfigure}{0.33\linewidth}
    \centering
    \includegraphics[width=1\linewidth]{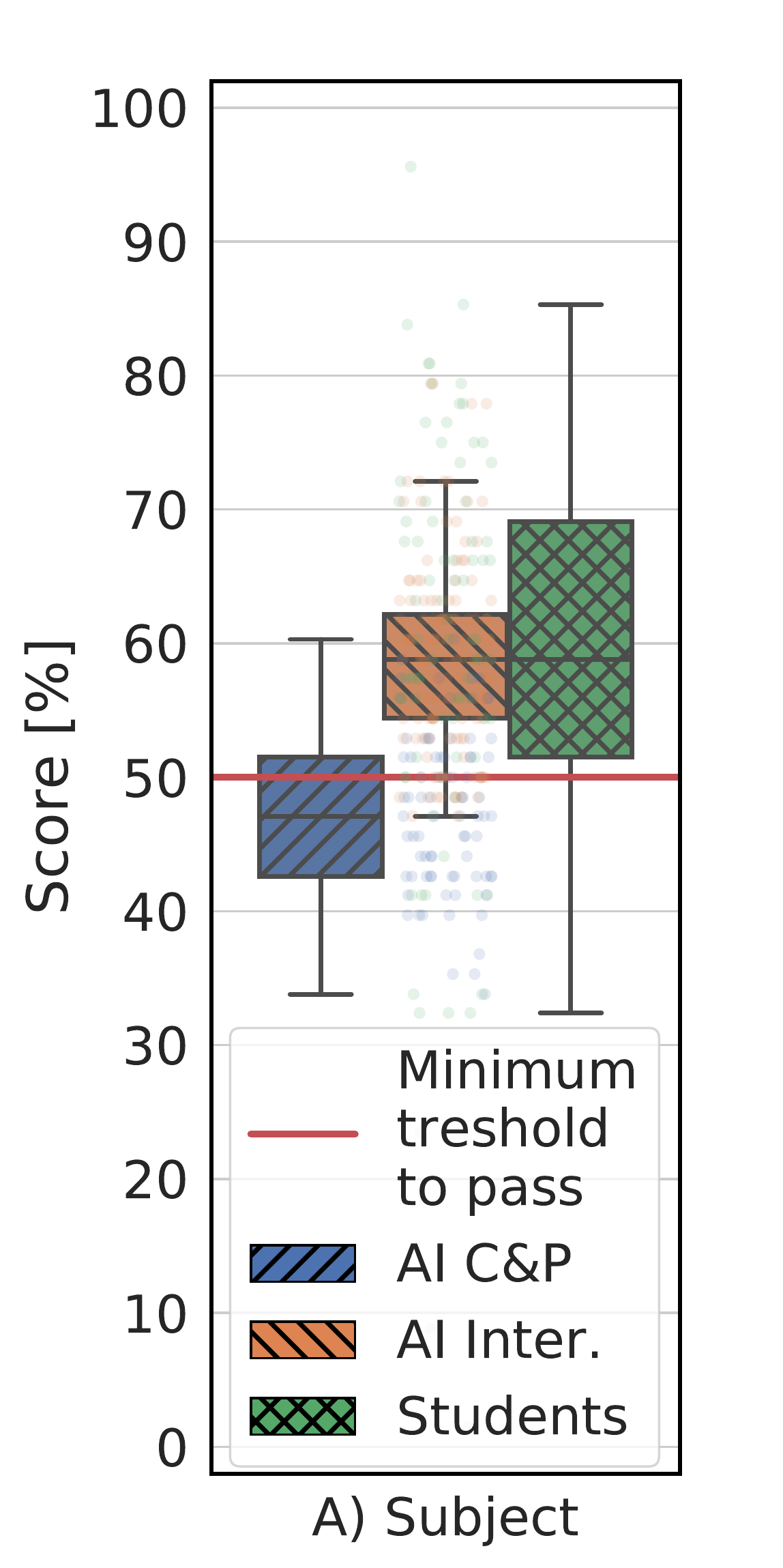}
	\label{aaa}
  \end{subfigure}%
  \begin{subfigure}{0.33\linewidth}
    \centering
    \includegraphics[width=1\linewidth]{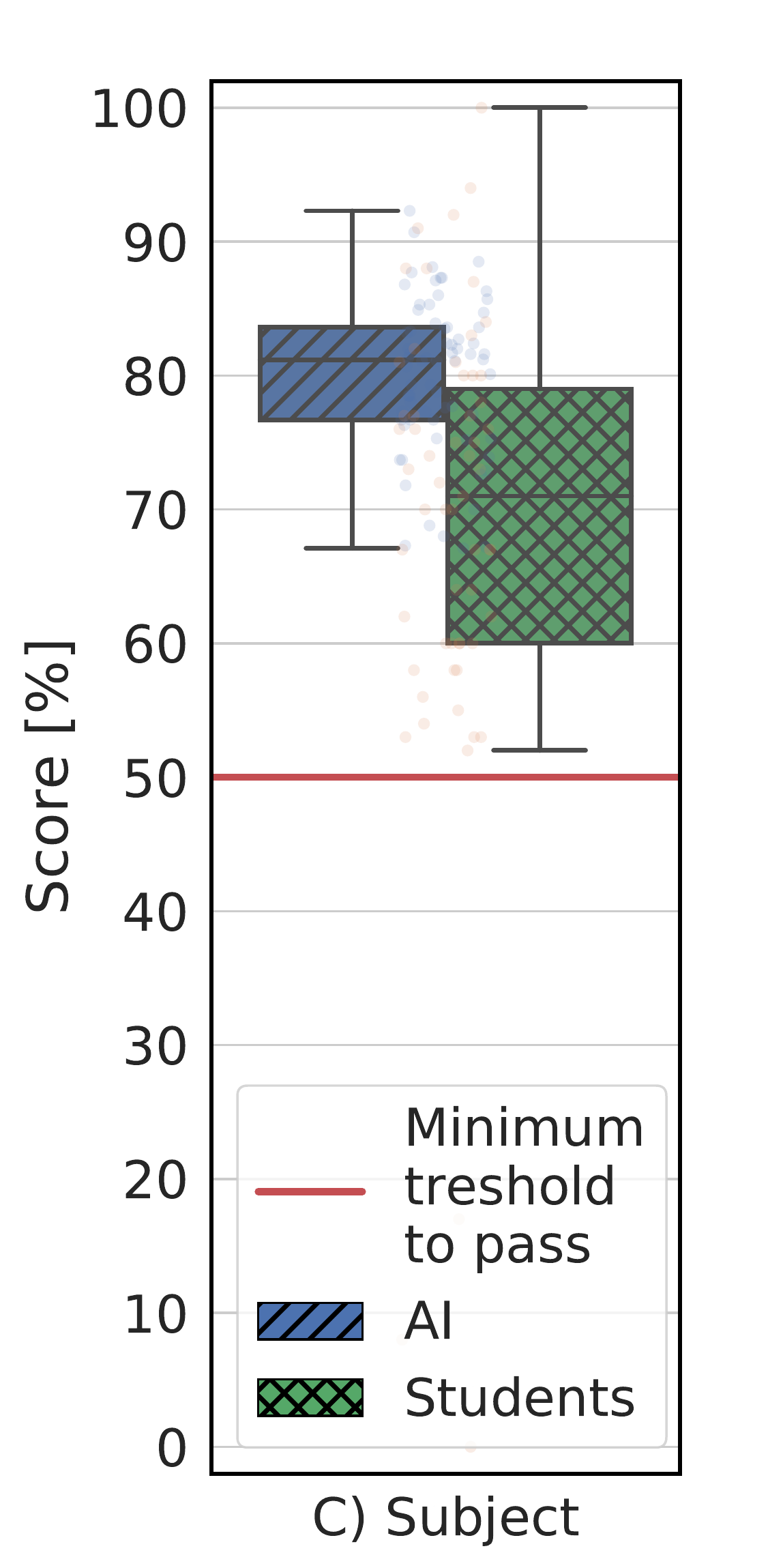}
    \label{bbb}
  \end{subfigure}
\begin{subfigure}{0.33\linewidth}
    \centering
    \includegraphics[width=1\linewidth]{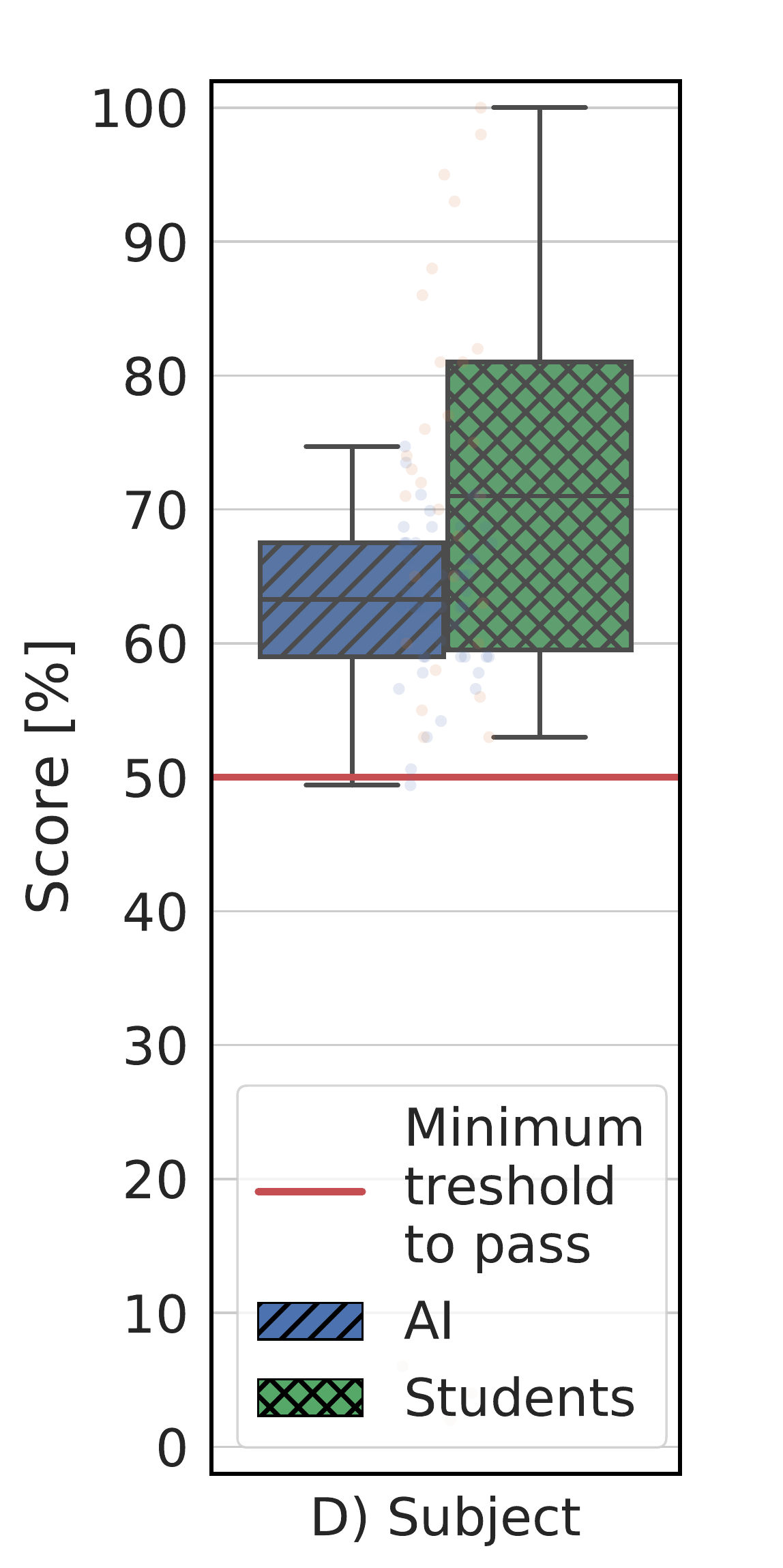}
    \label{ccc}
  \end{subfigure}
  \caption{Overall scoring assessment for the various subjects based on different examination methods with highlighted pass mark threshold.}
  \label{fig:assessment}
\end{figure}

\smallskip
\noindent
\emph{\textbf{Small Projects.}}
ChatGPT easily implemented algorithms such as Miller-Rabin, Solovay-Strassen, and Pollard-Rho.
Copying the output from ChatGPT significantly reduced the time and effort required.
However, this reduced the educational effect of the assignment as no study of the algorithms was needed.
A project completed in this way, in general, always received at least 40\% points or more, as can be seen in \autoref{fig:evaluation} [\texttt{C) Program.}].

\smallskip
\noindent
\emph{\textbf{Term Projects.}}
ChatGPT generated database schema from the assignment as an executable SQL script. The quality of the database design was of an average student according to the assessment made by three lectors. The proposed folder structure represented the MVC architecture, while most students neglected this.
ChatGPT easily generated code for database models, sample data, or views for login or registration pages. Interactive querying allows modifying the generated code to include styles or refactoring code. Moreover, the~generated code was tailored to fit the context of the initial database design (i.e., registration form fields).

\smallskip
\noindent
\emph{\textbf{Interactive Projects.}}
Caesar's cipher encrypted a secret message, which appears as the MOTD (Message Of The Day).
A simple request for AI to solve it failed; however, ChatGPT suggested frequency analysis and attempted to perform it, resulting in a bad outcome.
It eventually determined that the substitution cipher had been utilized.
AI attempts at decoding produced only incorrect results, similar to what we observed with exams, but at least gave us a correct hint on how to approach a given problem.

On the contrary, ChatGPT has successfully completed the easy task of finding a secret in a hidden file using Linux commands.
It also successfully solved a more advanced task of obtaining a~password from an obfuscated Javascript and reverting the SHA-1 hash.
ChatGPT identified the hash type and provided advice for revealing the password but was not allowed to help crack the hash (due to ethical constraints of the AI system).
Instead, it successfully directed towards online tools for revealing the original message.

ChatGPT cannot solve such an assignment independently.
It can only navigate students through the problem or demonstrate novel approaches or technologies.
In summary, ChatGPT slightly reduces the effort and knowledge needed to complete this assignment; however, most of the work remains to be done by students.

\subsubsection{\textbf{Final Assessment}}
\autoref{fig:assessment} portrays the final assessment results based on scored points for three courses.
To pass a course, a~student must earn at least 50\% of the total points available throughout the semester via various tasks, tests, and exams.
The evaluations for each course differed: Subject A had a test worth 60\% and a fulltext exam worth 40\% total points, Subject C had a programming assignment (30\%) and a fulltext exam (70\%), and Subject D had an essay (35\%) and a fulltext exam (65\%).
The key takeaway is that the ChatGPT did well overall and can pass all courses, except for a minor loss in Subject A when the AI was used in Copy\&Paste mode during the test examination technique.

\section{Discussion}
\label{sec:discussion}
The results indicate that the impact of ChatGPT usage by students is enormous. This section thus further discusses our findings, possible impacts, and mitigations.

\subsubsection{\textbf{Insights}}

During our experiments with ChatGPT, we observed numerous interesting insights. This section discusses the most important ones.

We observed great variability in the correctness of the ChatGPT answers. Several exams included an almost identical question on deciphering a message using Vigenère cipher. The answers were diametrally different for each query. In some cases, ChatGPT solved this task; in others, it provided completely wrong results. 

ChatGPT tends to make up events, links, and references. Several times, a link to a non-existing image, GitHub repository, or name of a publication was returned in the answer. If the author does not pay enough attention or does not understand the topic, such mistakes might be observed in the submitted text. This might serve as a good starting point for detecting AI-written papers.

\subsubsection{\textbf{Negative Impacts}}

ChatGPT allows for easy cheating in the university environment. Correct answers to exam questions, complete term papers, or functional pieces of code are available with a few clicks. This easy access to correct solutions can vastly reduce the learning experience that students should undergo.  

Using the ChatGPT to write term papers is, in fact, plagiarism. Moreover, replacing own writing with AI-generated text will stop original content creation. The same goes for programming. Using the chatbot to generate code vastly reduces the knowledge and experience students gain while solving the assigned problem. 

The ChatGPT is especially problematic for freshmen, as the learning process requires students to go from easy assignments to more challenging ones. As ChatGPT easily solves easy assignments without special intervention, students may not be required to learn the essential concepts and fail later at the more challenging assignments because of a lack of prior knowledge.
In contrast, ChatGPT sometimes tends to provide incorrect or misleading answers. This may lead to the propagation of misinformation if students accept these answers as a general truth.

The experiments executed in the Czech language strengthened our results. The threat model is not limited by the native language, as this tool might be used in an arbitrary language.
This might ultimately result in students passing the courses without properly understanding the substance. 

Finally, these impacts might be significantly amplified by, for example, another pandemic lockdown, where almost all interaction was solely through computers. Such a setting allows for very easy use of AI tools for cheating. 

\subsubsection{\textbf{Positive Impacts}}

Students are mostly required to work individually; however, in recent years, team projects and consulting individual projects with peers have become more common. This is a disadvantage for students who, for some reason, are not part of one of the social groups (social distancing, shame, introversion, \dots). This way, AI can pose a role of a teaching assistant. Students can discuss encountered problems or ideas for solving the assignment. Ultimately, the ChatGPT takes a role of an \emph{advanced rubber duck}~\cite{clinton_2022} that helps students and professionals solve encountered problems.

AI chatbots might also help to accelerate the learning process for IT-experienced individuals with new technologies. Well-prepared queries make the learning process more effective, as the chatbot can better explain the underlying concepts or provide sample codes in addition to the official documentation. This also applies to gaining more knowledge of computer science-related topics that the user is unfamiliar with.

\subsection{Are changes in education needed?}
In our opinion, yes.
Changes to the education model are inevitable, and the reaction should be immediate. The following areas should be addressed:

\subsubsection{\textbf{Detection}}
Tools for verifying the authenticity of submitted work should be used. Antiplagiarism or detection tools might be helpful in revealing AI-written tools. Especially in courses with many enrolled students, it will be challenging for educators to identify such works independently. We tested the GPTZero detector on the term papers. Our findings suggest that the detection is suited for the English language only. For English text, the received likelihood metrics sufficiently distinguished human and AI-written text. However, for text written in the Czech language, this difference vanished. It is thus questionable how to approach the detection in non-English speaking countries.

\subsubsection{\textbf{Prevention}}
The impacts of such technologies should be carefully discussed with students so they understand the potential impacts. Explaining the importance of academic integrity and the need to absorb basic concepts is crucial to allow the development of students into professionals. A suitable approach might be milestones during the study where students must demonstrate their ability to apply the knowledge in a controlled environment without the possibility of cheating. The assessment during these milestones will have the parameters of a real control mechanism. Not only for exams but also for programming assignments.

It is also possible to completely restrict access to ChatGPT, as was already done in New York~\cite{rosenblatt_2023}. As this restriction might be challenging to achieve during homework, several measures should be implemented to limit ChatGPT impacts.
One solution is to set the score of home-solved assignments to a minimum to have an insignificant impact on the final score.
Alternatively, specific libraries or feature limitations in programming tasks may make ChatGPT ineffective.  Moreover, any assignments or exam questions that require practical knowledge or problem-solving are not easily solved by ChatGPT, and a certain amount of knowledge and determination is still required. The same goes for schemas (images), as ChatGPT cannot produce or understand them.

\subsubsection{\textbf{Adoption}}
Educators and students must find a balance in using ChatGPT and related tools. As previously mentioned, the benefits associated with correct usage are not negligible. If students use this technology responsibly, their performance might be boosted. In addition, educators might benefit from the abilities of ChatGPT in curriculum creation or boosting their performance.

\section{Conclusions}
\label{sec:conclusions}

We demonstrated how easily ChatGPT could be misused and concluded that this AI might pass the courses required for a university degree. With a focus on IT security in general, we examined a~plethora of types of instruments to validate the students' knowledge.
This complex evaluation revealed that AI tools significantly threaten the integrity of the academic area. Without changes to the educational model, plagiarism and cheating will result in the production of low-quality graduates.

It can be expected that mere restrictions will not be sufficient or reasonable, not only because of the lack of detection tools and, thus, the inability to punish misuse. Instead, it will be necessary to find ways to prepare students for the effective use of this technology. Given the huge investments already announced, this technology will certainly be used in the future. Therefore, it is necessary to continue fostering desirable qualities in students, such as critical thinking, the ability to work without technology, or creativity.

However, we identified several benefits for students and educators. The major positive impacts are an AI assistant and the acceleration of the learning process.
The AI assistant is an excellent use case that diminishes the differences between students' performance based on their social groups. Now, every student has an assistant to help in discussing the encountered problems, which can even supplement the teacher in some ways. As a result, student performance might be boosted, and teachers' time is saved.

Moreover, this assistant can accelerate the learning process for new technologies. Experienced individuals can use ChatGPT to discuss and understand unknown technology without spending hours reading the documentation or experimenting. Ultimately, this might result in training the GPT model over documentation and specification of selected technology to create a kind of \emph{guru} that can help resolve most problems.

Finally, we expect many related works to be published soon, which will experimentally confirm our statements on the quality and usability of AI tools to be used for cheating. This will support the need for adequate changes and adaptations.

\begin{acks}
This work was supported by the internal project of Brno University of Technology FIT-S-23-8151.
\end{acks}

\bibliographystyle{bib/ACM-Reference-Format}
\bibliography{bib/references}

\appendix

\end{document}